# Dopants, Defects and Magnetism in Epitaxial $Co_xTi_{1-x}O_{2-x}$ Anatase

S.A. Chambers,[1] S.M. Heald,[1] R.F.C. Farrow,[2] J.-U. Thiele,[2] R.F. Marks,[2] M.F. Toney,[2] A. Chattopadhyay[2]

[1]*Fundamental Science Division, Pacific Northwest National Laboratory, Richland, Washington 99352*
[2]*IBM Almaden Research Center, San Jose, California 95120-6099*

We demonstrate that room-temperature ferromagnetism in epitaxial Co-doped $TiO_2$ anatase is driven by electron-mediated exchange interaction, and not by metallic Co clusters. Co(II) substitutes for Ti(VI) in the lattice and produces oxygen vacancies that do not contribute carriers. Free electrons originate with oxygen vacancies resulting from an oxygen deficiency during growth.

PACS numbers: 81.15.Hi, 75.50.Pp, 75.70.Ak

The quest for diluted magnetic semiconductors (DMS) which retain their magnetism at and above room temperature is spanning several classes of materials. Such materials are critically important in the development of spintronics as spin injectors for semiconductor heterostructures that can operate without cryogenic cooling. Group IV, III-V, and II-VI DMS materials typically exhibit Curie temperatures ($T_c$) well below ambient due to weak interaction of the magnetic impurities. Calculations based on the Zener model of magnetism suggest that the strongest interaction is that mediated by holes, and experimental studies carried out to date have borne out this prediction [1]. One notable exception is that of Mn-doped GaN, which grows *n*-type by gas-source molecular beam epitaxy under certain conditions, and appears to be ferromagnetic at room temperature [2]. In addition, it has recently been shown that at least one oxide semiconductor – Co-doped $TiO_2$ anatase or $Co_xTi_{1-x}O_{2-x}$ – is ferromagnetic well above room temperature when doped *n*-type by oxygen vacancies for x < ~0.1 [3-5], but the mechanism of magnetism remains unknown.

Here we show that magnetism occurs in this material because free electrons, generated by oxygen vacancies resulting from the growth conditions, ferromagnetically align the isolated Co(II) spins. Co(II) substitution for Ti(IV) in the anatase lattice results in oxygen vacancies, as dictated by charge neutrality requirements. However, these vacancies, which occur even under oxygen rich conditions, do not contribute to the conductivity. Thus, substitutional Co(II) *and* oxygen vacancies in excess of those required to compensate Co(II) must be present for ferromagnetism to occur. Dynamical mean-field calculations suggest that increasing the Co concentration results in a loss of magnetism as antiferromagnetic superexchange overtakes and dominates electron-mediated exchange interaction.

The growth of epitaxial $Co_xTi_{1-x}O_{2-x}$ on $LaAlO_3$(001) by oxygen-plasma-assisted molecular beam epitaxy has been described in detail elsewhere [6]. This growth method results in a majority (> ~95%) of phase-pure epitaxial anatase, along with minority phases of rutile, CoO and/or Ti-Co spinel, as verified by X-ray diffraction. None of these are ferromagnetic, at least in the bulk. Core-level x-ray photoemission measurements utilizing monochromatic Al Kα x-rays and a normal emission geometry were made *in situ*. Samples were then transferred through air to various experimental stations to measure conductivity, Kerr rotation, and Co K-shell x-ray absorption near-edge structure (XANES) and extended x-ray absorption fine structure (EXAFS). The latter two measurements were done at the PNC-CAT beamlines at the Advanced Photon Source at Argonne National Laboratory using a bend magnet and insertion device for parallel (*s* polarization) and perpendicular (*p* polarization) orientation of the electric field vector relative to the surface, respectively.

One of the critically important issues pertaining to this material concerns the chemical state of Co. Co present as segregated metal clusters would explain the robust magnetism without recourse to a DMS description. We show in Fig. 1 Co 2p spectra for a typical $Co_xTi_{1-x}O_{2-x}$ film, along with reference spectra for elemental Co, CoO, and metastable γ-$Co_2O_3$, all grown as pure films, the latter two also being epitaxial. In addition to different binding energies for the "primary" photoelectron peaks, there are large differences in the satellite peak structure for the different formal oxidation states of Co. The $Co_xTi_{1-x}O_{2-x}$ film spectrum matches the CoO reference spectrum very well, indicating that Co is in the +2 formal oxidation state in $Co_xTi_{1-x}O_{2-x}$. Significantly, there is no photoemission at the energy measured for Co metal, indicating the absence of this species near the surface of $Co_xTi_{1-x}O_{2-x}$. This observation has been made consistently for dozens of films with a wide range of Co concentrations. The probe depth of the Co 2p photoelectrons depicted in Fig. 1 is of the order of 50 to 60Å. Therefore, we cannot eliminate the possibility of elemental Co deeper in the film with these data. To do so, we have measured Co K-edge XANES for $Co_xTi_{1-x}O_{2-x}$ and have compared to reference spectra for a number of Co-containing compounds. The results are summarized in Fig. 2.

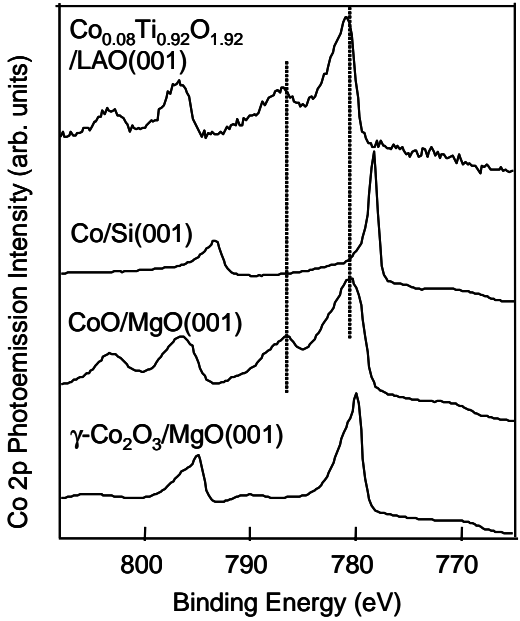

**FIG. 1.** *Co 2p spectra for 200Å of $Co_{0.08}Ti_{0.92}O_{1.92}$/ LaAlO₃(001), Co/Si(001), CoO/MgO(001), and γ-Co₂O₃/ MgO(001) obtained at normal emission.*

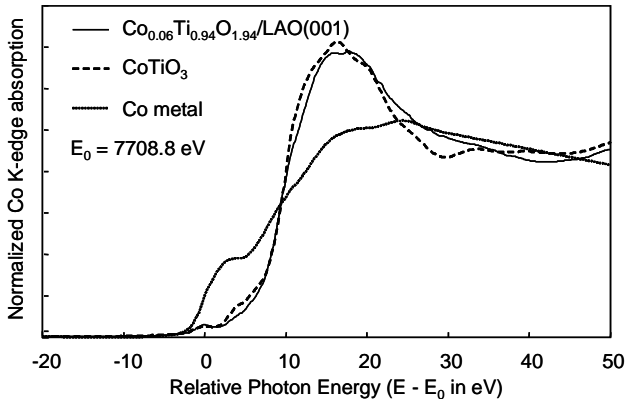

**FIG. 2.** *Co K-edge XANES for 200Å of $Co_{0.06}Ti_{0.94}O_{1.94}$ /LaAlO₃(001) obtained in p polarization, along with reference spectra for CoTiO₃ powder and Co metal foil. The spectra obtained in s polarization reveal negligible differences relative to those measured with p polarization.*

In addition to having a depth sensitivity that exceeds the film thickness, XANES has the advantage of high sensitivity to low concentrations of Co; submonolayer equivalents can be measured. There is a high degree of similarity between the $Co_xTi_{1-x}O_{2-x}$ film spectrum and that for CoTiO₃, again indicating a +2 formal oxidation state for Co in the film. Moreover, this strong similarity reveals that Co(II) in the film is in a local structural environment qualitatively similar to that of Co in CoTiO₃ – a distorted octahedral cage. This result suggests that Co(II) substitutes for Ti(IV) in the lattice, since the cations in anatase are also in a highly distorted octahedral field of oxygen ligands. The weak pre-edge resonance at $E - E_o = 0$ eV is due to a 1$s$ to 3$d$ transition which is strictly forbidden in octahedral coordination, but weakly allowed if the symmetry is broken. The observed magnitude is further evidence for distorted octahedral coordination. There is a large chemical shift between Co(0) and Co(II), as evidenced by the different threshold energies, making the detection of Co(0) straightforward. Clearly, there is no Co(0) in the $Co_{0.06}Ti_{0.94}O_{1.94}$ film. This result has been corroborated by x-ray diffraction and transmission electron microscopy measurements on other samples.

In order to determine the detailed structural environment of Co(II) in $Co_xTi_{1-x}O_{2-x}$, we turn to Co K-shell EXAFS, which is summarized in Fig. 3.

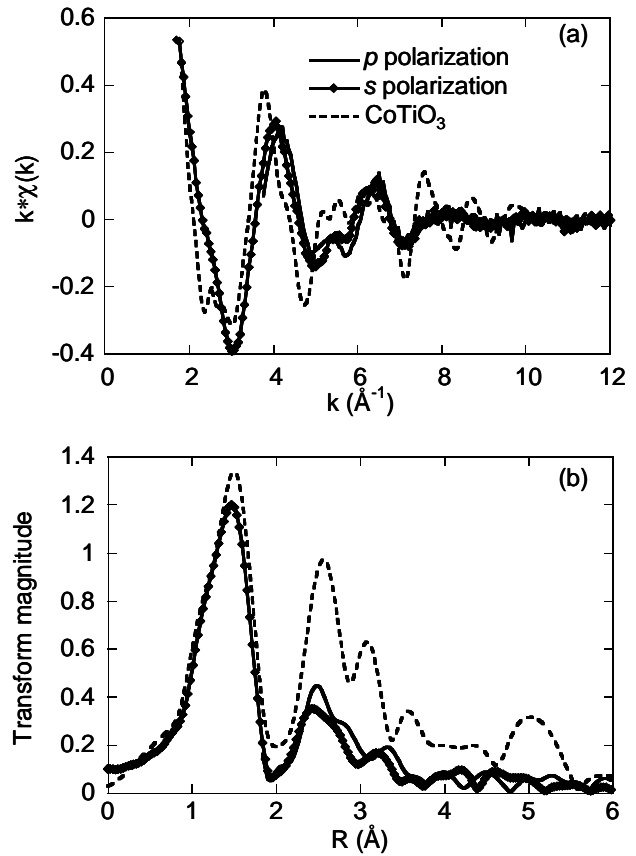

**FIG. 3.** *Co K-shell EXAFS for 200Å of $Co_{0.06}Ti_{0.94}O_{1.94}$/ LaAlO₃(001) measured using s and p polarization and compared to CoTiO₃ powder (a), and Fourier transforms of the same over the range k = 1.5-12Å⁻¹, with $k^2$ weighting (b).*

These data were analyzed using the Autobak and Ifeffit programs [7]. Model standards were calculated using Feff 7 [8]. The amplitudes were calibrated by fitting to CoO and CoTiO₃. Both gave identical amplitude reduction factors of 0.80. Simultaneously fitting the data obtained using $s$ and $p$ polarizations yields the following structural parameters: $R_{ab}$ = 2.04 +/- 0.01Å and $\sigma^2_{ab}$ = 0.0096, $R_c$ = 2.05 +/- 0.01Å and $\sigma^2_c$ = 0.011, and $N$ = 5.45 +/- 0.3. Here, $R_{ab}$ ($\sigma^2_{ab}$) and $R_c$ ($\sigma^2_c$) are the Co-O bond lengths (Debye-Waller factors) in the $ab$ plane and along the $c$ direction, respectively, and $N$ is the spherically-averaged effective coordination number for the Co. These bond lengths are significantly larger than those found in pure anatase (1.94Å (4x) and 1.97Å (2x)), as expected in light of the larger Co-O bond lengths in CoTiO$_3$ (2.08Å (3x) and 2.20Å (3x)). Thus, local strain fields are surely present at substitution sites. In addition, the Debye-Waller factors are higher than expected for Co-O bonds at room temperature [9], the most probable cause being the presence of structural disorder at Co sites resulting from strain and thin-film growth. Finally, $N$ is less than 6.00, the value expected for perfect octahedral coordination.

The strain introduced into the lattice by Co substitution for Ti may explain the facts that: (i) Co is in the +2 formal oxidation state, and, (ii) $N$ is less than 6. Strain relief can be achieved by expelling oxygen from the lattice. The formal charge on a Co will be lowered if adjacent oxygen vacancies exist. In point of fact, one oxygen vacancy must be created for every Co(II) that substitutes for Ti(IV) in the lattice in order to maintain charge neutrality. Thus, the correct empirical formula for Co-doped TiO$_2$ is Co$_x$Ti$_{1-x}$O$_{2-x}$. A value of $N$ equal to 5.00 is expected if these vacancies are perfectly correlated with substitutional Co(II). If, on the other hand, these vacancies are randomly distributed throughout the lattice, $N$ will equal 5.82 for x = 0.06 [10]. The fact that $N$ assumes an intermediate value reveals significant, but incomplete structural correlation between oxygen vacancies and substitutional Co(II). Significantly, we have found that oxygen vacancies associated with substitutional Co do not contribute to the electrical conductivity of the material. We have grown films with a few to several at. % Co that are insulating. Conductivity in these films is directly correlated with the oxygen pressure and plasma power level used during growth. We have found a monatomic relationship between film resistivity and oxygen pressure and/or plasma power level for pure TiO$_2$ anatase growth [6]. It has been shown that oxygen vacancies in anatase produce shallow donor levels that dope the material $n$ type [11]. We have found empirically that oxygen vacancies in excess of those needed to compensate substitutional Co(II) in Co$_x$Ti$_{1-x}$O$_{2-x}$ anatase are required to make the material semiconducting.

In addition, we have found an inverse relationship between the room-temperature Kerr rotation and film resistivity. We show in Fig. 4 the dependence of the polar Kerr rotation, which is directly proportional to the out-of-plane magnetization, on film resistivity for a wide range of Co mole fractions. Note, however, that at the wavelength of the He/Ne laser used here, λ = 632.8 nm or hν = 1.96 eV, the spectral dependence of the Kerr rotation is very close to a zero crossing [12]. Therefore, e.g., for films grown with significantly higher deposition rates, the spectral features of the Kerr rotation may shift, resulting in larger positive, or under some conditions, negative Kerr rotation. For the films shown here, all grown under identical conditions, the Kerr rotation decreases with increasing x, but also depends critically on the carrier concentration, as expected for a DMS. Films with x > ~0.1 are not magnetic under any circumstance, as verified for a number of films by vibrating sample magnetometry. These results have to do with the nature of the magnetic sublattice in this material. Increasing x reduces the mean separation between Co ions such that an antiferromagnetic superexchange interaction overtakes electron-mediated exchange interaction, thereby eliminating the ferromagnetism. The underlying structure of Co$_x$Ti$_{1-x}$O$_{2-x}$ is that of a Kondo lattice where a fraction x of the cation sites are magnetic and the remainder (1-x) are non-magnetic. At each of the Co sites, there is a local ferromagnetic interaction that couples a core spin to an itinerant electron. This model is quite similar to the binary alloy problem, and we can solve the problem using dynamical mean field theory [13,14]. If we denote the mean-field function at the Co sites as g(ω), then the following self-consistency condition emerges: g(ω) = ω + μ - $t^2$G$_{avg}$[g(ω)]. Here ω, μ and t are frequency, chemical potential and the hopping parameter respectively, and the average Green's function is given by an algebraic average of the Green's functions at interacting Co and non-interacting Ti sites.

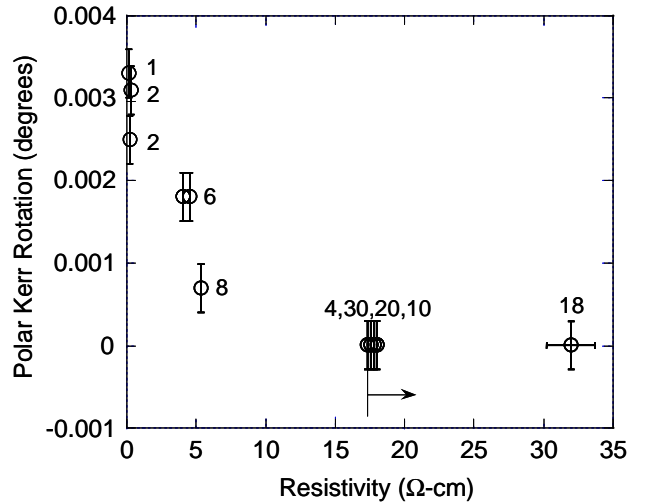

**Fig. 4.** *Polar Kerr rotation vs. film resistivity for several 200Å-thick Co$_x$Ti$_{1-x}$O$_{2-x}$ films on LaAlO$_3$(001). The numbers adjacent to the data points are the percents of Co in the cation sublattice. The 18 Ω-cm value is the lower resistivity limit for the locus of points centered there, and the resistivity error bar is large for the point at 32 Ω-cm because of the low current value. The uncertainty in the other resistivities is the diameter of the circle.*

This calculation provides a qualitative explanation for the nature of Fig. 4. A central feature of this plot is that the films become less magnetic as well as more resistive with increasing Co concentration. This behavior evolves naturally in the ferromagnetic Kondo lattice that we have investigated theoretically. For very dilute Co ($x < 0.01$), the ground state of the lattice is a ferromagnet with an ordering wave vector $\mathbf{q} = (0,0,0)$. For such electron-mediated ferromagnets, the metallicity is closely tied to the system being ferromagnetic. Having neighboring spins ferromagnetically aligned provides the maximum probability for the carrier hopping from one magnetic site to another, thereby making it a metal. As the Co concentration is increased, the energy of some magnetically incommensurate state of finite $\mathbf{q}$ becomes very close to that of the ferromagnetic state and eventually becomes lower than the $\mathbf{q}=0$ state, resulting in the system ordering at some incommensurate wave vector. Thus, the ferromagnetism is reduced as the Co concentration is increased. This phenomenon also results in reduced carrier hopping and an associated increase in resistivity. The transition between the ferromagnetic state and the various incommensurate magnetic states is second order. With increasing Co, the system goes through a series of such non-zero $\mathbf{q}$ states with appropriately decreasing magnetization and conductivity. For the state where there is Co substitution at every Ti site, the ground state is a commensurate antiferromagnet with $\mathbf{q} = (\pi, \pi, \pi)$. This is known experimentally as well; CoO is indeed an antiferromagnet. The transition from a state of finite $\mathbf{q}$ to the antiferromagnetic state is seemingly first-order, which may give rise to magnetic phase separation in the sample.

This theory is of qualitative value in determining $T_c$ for $Co_xTi_{1-x}O_{2-x}$. Our best estimates from dynamical mean-field theory, assuming 3-fold degenerate Co $t_{2g}$ states, result in a $T_c$ value of ~500K, which is somewhat lower than the current experimental estimate (~700K). This situation will surely improve with further input from first-principles calculations.

In summary, we have shown experimentally that Co substitutes for Ti(VI) in the $Co_xTi_{1-x}O_{2-x}$ anatase lattice, and that oxygen vacancy defects accompany this substitution in order to relieve strain. Co assumes a +2 formal oxidation state as a result. These vacancies are not electrically active, and do not make the material semiconducting. Additional oxygen vacancies, resulting from a slight oxygen under-pressure during growth, generate itinerate electrons and enhance both the conductivity and the magnetism to an extent that is inversely proportional to the oxygen pressure and plasma power. These free carriers act to couple the Co spins in a way that is at least qualitatively consistent with theoretical predictions based on a Kondo lattice model. $Co_xTi_{1-x}O_{2-x}$ thus behaves as expected for an *n*-type DMS. That $Co_xTi_{1-x}O_{2-x}$ is *n*-type, and exhibits a $T_c$ value well above room temperature, bodes well of it's use as a spin injector for semiconductor heterostructures that can operate without cryogenic cooling.

The film growth and *in-situ* materials characterization described in this paper were performed in the

Environmental Molecular Sciences Laboratory, a national scientific user facility sponsored by the Department of Energy's Office of Biological and Environmental Research and located at Pacific Northwest National Laboratory. This work was supported by the PNNL Nanoscience and Technology Initiative and the US Department of Energy, Office of Science, Office of Basic Energy Sciences, Division of Materials Science. The work at IBM was supported by DARPA (grant no. DAAD 19-01-C-0060). PNC-CAT facilities and research at these facilities are supported by the US DOE Office of Science grant no. DE-FG03-97ER45628

____________________

[1] T. Dietl, H. Ohno, F. Matsukura, J. Cibert, D. Ferrand, Science **287**, 1019 (2000).
[2] G.T. Thaler, M.E. Overberg, B. Gila, R. Frazier, C.R. Abernathy, S.J. Pearton, J.S. Lee, S.Y. Lee, Y.D. Park, Z.G. Khim, J. Kim, F. Ren, Appl. Phys. Lett. **80**, 3964 (2002).
[3] Y. Matsumoto, M. Murakami, T. Shono, T. Hasegawa, T. Fukumura, M. Kawasaki, P, Ahmet, T. Chikyow, S.-Y. Koshihara, and H. Koinuma, Science, **291**, 854 (2001).
[4] S.A. Chambers, S. Thevuthasan, R.F.C. Farrow, R.F. Marks, J.-U. Thiele, L. Folks, M.G. Samant, A.J. Kellock, N. Ruzycki, D.L. Ederer, U. Diebold, Appl. Phys. Lett. **79**, 3467 (2001).
[5] S.A. Chambers, Mat. Today, April issue, p. 34 (2002).
[6] S.A. Chambers, C.M. Wang, S. Thevuthasan, T. Droubay, D.E. McCready, A.S. Lea, V. Shutthanandan, and C.F. Windisch, Jr., Thin Solid Films, to appear (2002).
[7] M. Newville, J. Synchrotron Rad. **8**, 322 (2001).
[8] S.I. Zabinsky, J.J. Rehr, A. Ankudinov, R.C. Albers and M.J. Eller, Phys. Rev. **B52**, 2995 (1995).
[9] The measured values for the CoO and $CoTiO_3$ standards range from 0.006 to 0.008.
[10] The value of x for this particular film was 0.06. There were thus 3% oxygen vacancies which if randomly distributed among six O ligands, would result in $N = 0.97*6 = 5.82$.
[11] L. Forro, O. Chauvet, D. Emin, L. Zuppiroli, H. Berger, F. Levy, J. Appl. Phys. **75**, 633 (1994).
[12] P. Fumagalli (unpublished).
[13] A. Chattopadhyay, S. Das Sarma and A. J. Millis, Phys. Rev. Lett. **87**, 227202 (2001).
[14] A. Chattopadhyay, A. J. Millis and S. Das Sarma, Phys. Rev. **B64,** 012416 (2001).